\documentclass[conference]{IEEEtran}
\IEEEoverridecommandlockouts

\usepackage{cite}
\usepackage{amsmath,amssymb,amsfonts}
\usepackage{algorithmic}
\usepackage{graphicx}
\usepackage{textcomp}
\usepackage{xcolor}
\usepackage{algorithm}

\usepackage{amsthm}
\usepackage{booktabs}

\theoremstyle{definition}
\newtheorem{definition}{Definition}

\usepackage{subcaption}

\def\BibTeX{{\rm B\kern-.05em{\sc i\kern-.025em b}\kern-.08em
    T\kern-.1667em\lower.7ex\hbox{E}\kern-.125emX}}
\begin{document}

\title{5G LDPC Codes as Root LDPC Codes via Diversity Alignment \\
}

\author{
Hyuntae Ahn\IEEEauthorrefmark{1},
Inki Kim\IEEEauthorrefmark{1},
Hee-Youl Kwak\IEEEauthorrefmark{2},
Yongjune Kim\IEEEauthorrefmark{3},
Chanki Kim\IEEEauthorrefmark{4}
and 
Sang-Hyo Kim\IEEEauthorrefmark{1}
\\[0.3em]
\IEEEauthorrefmark{1}Dept. of Electrical and Computer Engineering, Sungkyunkwan University, Suwon, South Korea\\
Email: \{dks1408@g.skku.edu, inkikim37@gmail.com, iamshkim@skku.edu\}\\[0.2em]
\IEEEauthorrefmark{2} Dept. of Electrical, Electronic and Computer Engineering, University of Ulsan, Ulsan, South Korea\\
Email: ghy1228@gmail.com\\[0.2em]
\IEEEauthorrefmark{3} Dept. of Electrical Engineering, POSTECH, Pohang, South Korea\\
Email: yongjune@postech.ac.kr\\[0.2em]
\IEEEauthorrefmark{4} Dept. of Computer Science and Artificial Intelligence, Jeonbuk National University, Jeonju, South Korea\\
Email: carisis@jbnu.ac.kr\\[0.2em]
}

\maketitle

\begin{abstract}
This paper studies the diversity of protograph-based quasi-cyclic low-density parity-check (QC-LDPC) codes over nonergodic block-fading channels under iterative belief-propagation decoding. We introduce diversity evolution (DivE), a Boolean-function-based analysis method that tracks how the fading dependence of belief-propagation  messages evolves across decoding iterations. Under a Boolean approximation of block fading, DivE derives a Boolean fading function for each variable node (VN) output (i.e., the a-posteriori reliability after iterative decoding), from which the VN diversity order can be directly determined. Building on this insight, we develop a greedy block-mapping search that assigns protograph VNs to fading blocks so that all information VNs achieve full diversity, while including the minimum additional parity VNs when full diversity is infeasible at the nominal rate. Numerical results on the 5G New Radio LDPC codes show that the proposed search finds block mappings that guarantee full diversity for all information bits without modifying the base-graph structure, yielding a markedly steeper high-SNR slope and lower BLER than random mappings.
\end{abstract}


%Introduction
\section{Introduction}
\label{sec:introduction}

Low-density parity-check (LDPC) codes have evolved from a theoretical cornerstone of capacity-approaching theory to a practical standard for 5th Generation-New Radio (5G-NR), favored for their hardware-friendly protograph-based structures~\cite{Gallager1962,Richardson2001a,3GPP38212_2020,Thorpe2003,Myung2005}.

However, while extensively optimized for additive white Gaussian noise channels (AWGNCs), these codes often encounter nonergodic block-fading channels (BFCs) where performance is governed by diversity order rather than a traditional noise threshold. In such regimes, the achievable diversity is fundamentally limited by a Singleton-like bound~\cite{Malkamaki1999,Fabregas2006}.

Root-LDPC codes \cite{Boutros2010,Kim20} were introduced to guarantee full diversity via specific ``rootcheck" constraints; however, their structural rigidity compromises the AWGN performance. This conflict raises a critical question: Can AWGN-optimized codes such as 5G-NR codes achieve full diversity without modifying their original structure under BFC?

Inspired by the diversity alignment (DA) framework that achieves simultaneous optimality over both AWGNCs and BFCs using a single polar code~\cite{Ju2022,Ju2025}. We seek to develop a coding scheme that attains full diversity using AWGN-optimized LDPC codes.
To this end, we first establish an analytical framework termed diversity evolution (DivE) similar to one in \cite{Ju2025}, which leverages Boolean approximations of fading to trace the evolution of diversity behavior of the edge messages across the iterative belief-propagation (BP) decoding. 
Using DivE, we can evaluate the diversity order of the variable nodes (VNs) in the protograph under iterative BP decoding  for an arbitrary block mapping. 
While DivE can also guide joint protograph design for BFCs and AWGNCs~\cite{Inki}, we focus on enhancing the BFC performance of existing AWGN-optimized (standard) LDPC codes via block-mapping optimization.

DivE enables an optimization of block mappings through a greedy search approach. Specifically, the block mapping is determined through an incremental assignment process, where DivE is performed at each step to analyze the impact of mapping choices on the resulting diversity. By evaluating the diversity contribution of each assignment, we can systematically select block indices that maximize the number of VNs achieving full diversity. 
Using this search, a near diversity-optimal block mapping can be found for 5G-NR LDPC codes, which can be called a root LDPC codes in a broader sense.

%%%%%%%%%%%%%%%%%%%%%%%%%%%%%%%%%%%%%%%%%%%%%%%%%%%%%%%%%%%%%%%%%%%
%Preliminaries
\section{Preliminaries}
\label{sec:prelim}

\subsection{Block-Fading Channels}
We consider BFC where a codeword is transmitted over $M$ independent fading blocks \cite{Knopp2000,Boutros2010}.
Let $N$ denote the codeword length, and assume that $N/M$ coded bits are transmitted over each fading block \cite{Knopp2000}.

Let $\mathbf{c}=(c_0,c_1,\ldots,c_{N-1})\in\{0,1\}^N$ denote the transmitted codeword.
We consider binary phase shift keying (BPSK) modulation, i.e., $s_i = 1-2c_i$, while the proposed framework can be extended to general $M$-ary modulations.
The received signal of the $m$-th fading block is given by
$
\mathbf{r}_m = h_m \mathbf{s}_m + \mathbf{n}_m, \qquad m=0, \ldots ,M-1
$
where $\mathbf{s}_m$ denotes the transmitted symbol vector assigned to the $m$-th block, $\mathbf{r}_m$ is the corresponding received vector, $\mathbf{n}_m$ denotes AWGN vector, and $h_m$ is the corresponding fading coefficient which follows $\mathcal{CN}(0,1)$, complex Gaussian with zero mean and unit variance \cite{Knopp2000,Boutros2010}.

The diversity order of a coded system is defined as
\begin{equation}
d_c \triangleq -\lim_{\gamma\to\infty}\frac{\log P_e(\gamma)}{\log{\gamma}},
\end{equation}
where $P_e(\gamma)$ denotes a block error rate (BLER) at signal-to-noise ratio (SNR) $\gamma$ \cite{Malkamaki1999,Fabregas2006}.

For the $M$-block setting, the achievable diversity order is fundamentally limited by a Singleton-like bound.
In particular, for code rate $R$, the diversity order $d_c$ satisfies
$
d_c \le 1+\left\lfloor M(1-R)\right\rfloor
\label{eq_singleton}
$ called a Singleton-like bound, 
where and $d_c=M$ is referred to as \emph{full diversity} \cite{Fabregas2006}.

\subsection{LDPC Codes and BP Decoding}
An LDPC code is specified by a sparse parity-check matrix $\mathbf{H}\in\mathbb{F}_2^{(N-K)\times N}$ and its Tanner graph, which consists of VNs and check nodes (CNs) \cite{Gallager1962,Tanner1981}.
Let $\mathcal{N}_v(i)$ denote the set of CNs connected to VN $V_i$, and let $\mathcal{N}_c(j)$ denote the set of VNs connected to CN $C_j$.
We consider iterative BP decoding in the log-likelihood ratio (LLR) domain \cite{Richardson2001b,Kschischang2001}.
We analyze the performance with the min-sum decoder (MSD) as a performance lower bound of BP decoder as in \cite{Boutros2010,Ju2025}. Let $L_{ch}(i)$ be the channel LLR for VN $V_i$, and let $L_{ji}$ and $R_{ji}$ be the CN-to-VN and VN-to-CN messages, respectively.
Under the min-sum update, the message updates are given by \cite{Fossorier1999,ChenFossorier2002,Richardson2001b}.
\begin{align}
    L_{ji} &= \left(\prod_{i'\in\mathcal{N}_c(j)\setminus\{i\}} \mathrm{sign}\!\left(R_{ji'}\right)\right) \cdot \min_{i'\in\mathcal{N}_c(j)\setminus\{i\}} \left|R_{ji'}\right|, \label{eq:cn_update} \\
    R_{ji} &= L_{ch}(i) + \sum_{j'\in\mathcal{N}_v(i)\setminus\{j\}} L_{j'i}. \label{eq:vn_update}
\end{align}
After a given number of iterations, hard decisions are made from the a-posteriori LLRs.

\begin{figure}[t]
    \centering
    \includegraphics[width=0.55\linewidth]{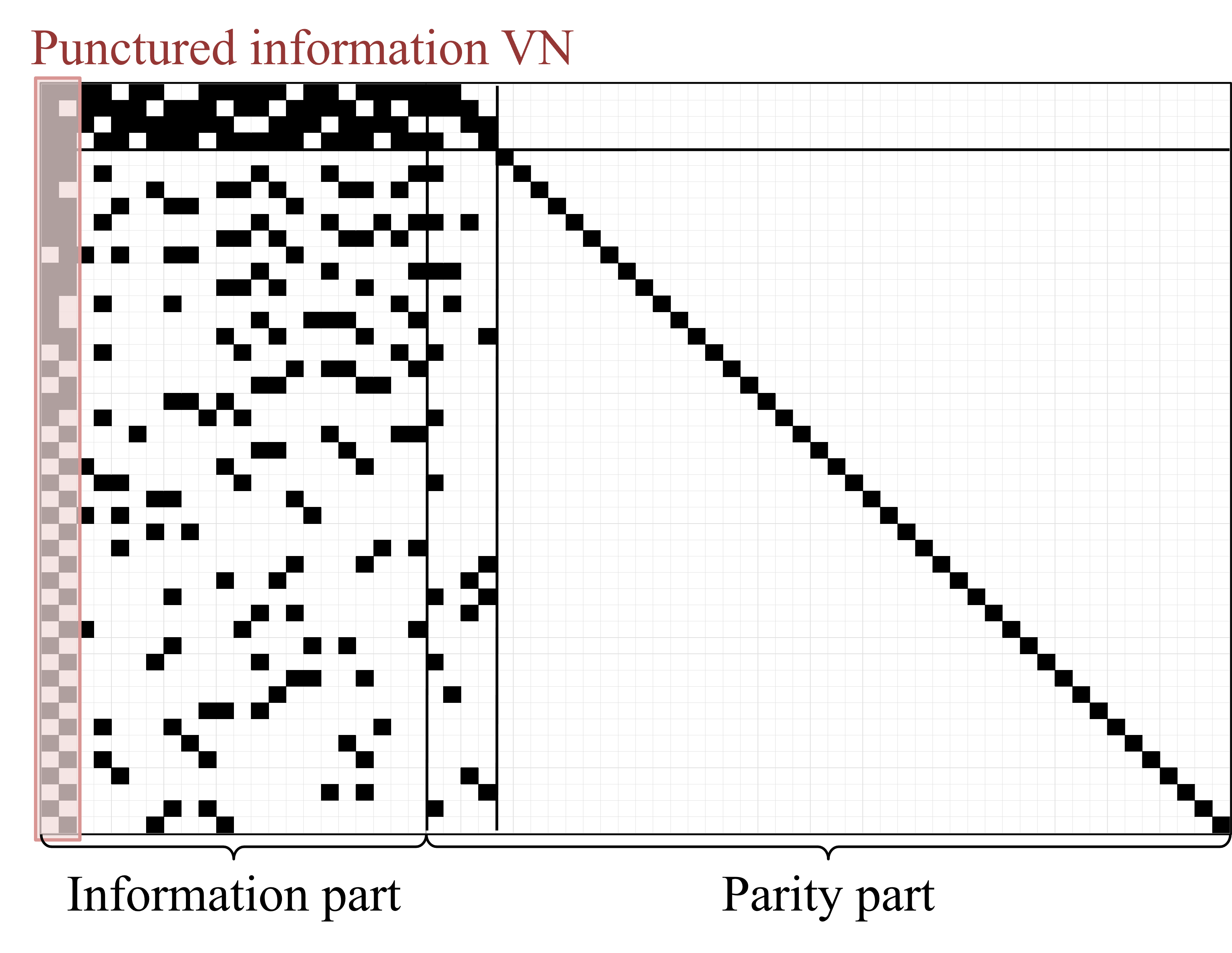}
    \caption{Base matrix of BG1 5G-NR LDPC codes: Black squares indicate the edges in the corresponding protograph. } 
    \label{fig:BG1}
\end{figure}

\subsection{Protograph-based and 5G-NR LDPC Codes}
We consider protograph-based QC-LDPC codes \cite{Thorpe2003}.
A protograph is a small bipartite graph $\mathbb{G}_p(\mathcal{V},\mathcal{C},\mathcal{E})$, where $\mathcal{V}$ and $\mathcal{C}$ denote the sets of VNs and CNs, respectively, and $\mathcal{E}$ denotes the set of edges.
It is represented by a base matrix $\mathbf{H}_p\in\mathbb{Z}_{\ge0}^{(n-k)\times n}$ whose entry $[\mathbf{H}_p]_{j,i}$ equals the number of edges between the $i$-th VN and the $j$-th CN \cite{Thorpe2003}.
Let $n$ denote the number of protograph VNs (columns of $\mathbf{H}_p$) and let $n-k$ denote the number of protograph CNs (rows of $\mathbf{H}_p$).
Accordingly, $k$ denotes the number of information protograph VNs.

A finite-length QC-LDPC PCM $\mathbf{H}\in\mathbb{F}_2^{(N-K)\times N}$ is obtained by lifting with factor $Z$, and lifting by $Z$ yields $N=nZ$ and $K=kZ$.\cite{Thorpe2003,3GPP38212_2020}.

We use this 5G-NR LDPC codes as a main example of protograph LDPC codes. In 5G-NR, data-channel coding employs two protographs (base graphs) BG1 and BG2 \cite{3GPP38212_2020}. The structure of BG1 is given  in Fig.~\ref{fig:BG1}, where the first two VNs are punctured as specified in~\cite{3GPP38212_2020}.

\subsection{Root LDPC Codes}
Root-LDPC codes were introduced in~\cite{Boutros2010} as LDPC constructions tailored to nonergodic BFCs.
In~\cite{Boutros2010}, root-LDPC codes are proposed by designing LDPC Tanner graphs that achieve diversity-optimal performance over BFCs under BP decoding.

This is achieved by introducing \emph{rootchecks}, specially structured CNs that are composed of a target information VN and other VNs. As in Fig.~\ref{fig:con_rootcheck}, a rootcheck is composed of one VN transmitted over a $m$-th fading block and all remaining VNs  transmitted over a common $m'$-th fading block different from $m$, $m'\neq m$.  
Through the rootcheck structure, the CN provides diversity 2 to the target VN. 

%%%%%%%%%%%%%%%%%%%%%%%%%%%%%%%%%%%%%%%%%%%%%%%%%%%%%%%%%%%%%%%%%%%
%Boolean expression
\section{Diversity analysis of LDPC Codes in Block Fading Channels}
\label{sec:boolean_dive}

First, we aim to evaluate of the decoding diversity of each VN and the entire code under iterative MSD for a given block mapping, which is a mapping from VNs to fading blocks. 
Inspired by \cite{Ju2025}, we develop the diversity evolution method for LDPC codes, leveraging a Boolean approximation of the fading (\textit{e.g.}, Rayleigh).

\begin{figure}[t]
\centering
\begin{minipage}{0.23\textwidth}
    \centering
    \includegraphics[width=0.8\textwidth]{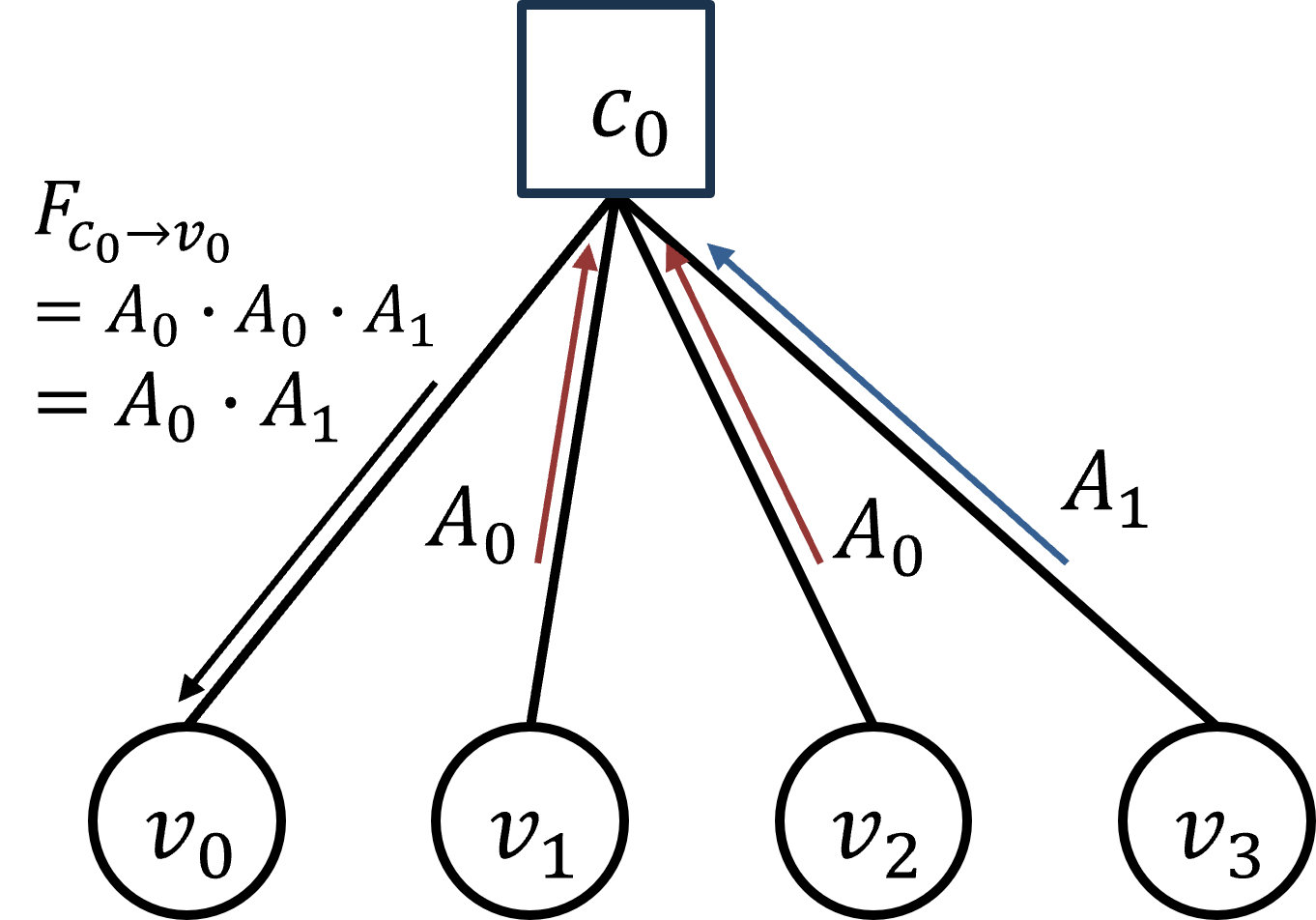}
    \subcaption{ } 
    \label{fig:CN_update}
\end{minipage}
\hfill
\begin{minipage}{0.23\textwidth}
    \centering
    \includegraphics[width=0.8\textwidth]{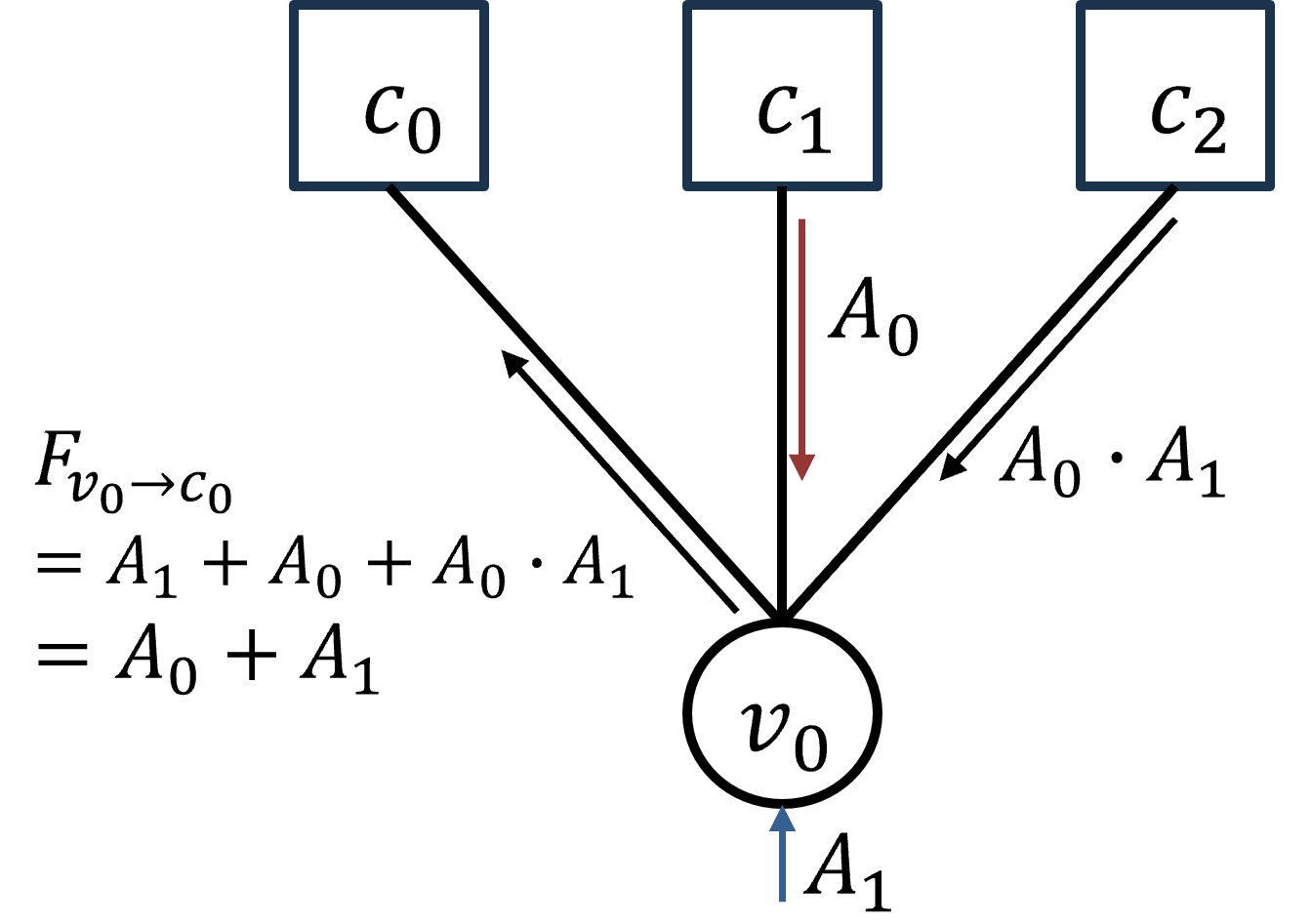}
    \subcaption{ }
    \label{fig:VN_update}
\end{minipage}
\caption{CN/VN processing of DivE (update of algebraic expression): (a) CN update, (b) VN update.}
\label{fig:Boolean_function_update}
\end{figure}

\subsection{DivE: Fading Analysis of MSD via Boolean Approximation}
\label{subsec:boolean_fading_msd}

We adopt a Boolean approximation $A_m$ of the fading coefficient $h_m$ for each block $m\in\{0,\ldots,M-1\}$ by the indicator of the fading and nonfading state as $A_m = \mathbf{1}_{|h_m|^2 \gamma \geq \rho_0}$ where $\rho_0$ is a threshold and $\mathbf{1}_E$ is the indicator function for event $E$ \cite{Ju2022,Ju2025}.
Let $\mathbf{A}=(A_0,\ldots,A_{M-1})\in\{0,1\}^M$ be the approximate fading where $A_m$ is the fading state of the $m$
-th block. Along with the iterative MSD process in \eqref{eq:cn_update} and \eqref{eq:vn_update}, the fading states of edge messages (of the MSD) are computed via symbolic product (Boolean AND) and addition (Boolean OR) operations following the framework in \cite{Ju2025} (see  Fig.~\ref{fig:Boolean_function_update}). This procedure constitutes a message passing algorithm over the protograph where the messages are Boolean algebraic expressions rather than scalar values, thereby tracking the diversity behavior. We refer to this process as \textit{diversity evolution (DivE)} for LDPC codes. 

\begin{algorithm}[t]
\caption{Diversity evolution over protograph}
\label{alg:DivE}
\begin{algorithmic}[1]
\footnotesize
\REQUIRE Graph  $\mathbb{G}_p,$ fading $\mathbf{A},$ block mapping $ {\pi(\cdot)}\in\{0,\ldots,M-1\}^n$. 
\ENSURE Fading functions of VNs

\STATE \textit{// 1. Initialization}
\STATE Assign $A_{\pi(i)}$ to the $i$-th VN for all $i$. 
\STATE \textit{// 2. Main loop}
\FOR{$0\leq \ell\leq$ max number of iterations}
    \STATE Check node update for all CNs as in Fig.~\ref{fig:CN_update}.
    \STATE Variable node update for all VNs as in Fig.~\ref{fig:VN_update}. 
\ENDFOR
\STATE \textit{// 3. Output fading function}
\STATE Output fading function update for all VNs (addition of all incoming messages.) 
\end{algorithmic}
\end{algorithm}

To specify the procedure, we formally define the block mapping, the mapping from VNs to fading blocks as: 
\begin{definition}[Block mapping]
\label{def:block_mapping}
Let $\pi:\{0,\ldots, n-1\}\to\{0,\ldots,M-1\}$ be a mapping that assigns the VN index $i$ to the fading-block index $\pi(i)$
Given a fading realization $\mathbf{A}$, the Boolean channel state associated with VN $v_i$ is given by $A_{\pi(i)}$.
\end{definition}
Then we can describe the DivE as in Algorithm~\ref{alg:DivE}. 
The output of DivE is the vector of Boolean functions whose entries are corresponding to VNs, respectively. 
Using DivE, it is easy to verify that existing root LDPC codes \cite{Boutros2010,Kim20} satisfy the optimal diversity.

The code achieves full diversity if all information VNs have full diversity. 
For the case of $M=2$, the functional update is rather straightforward as the number of all possible Boolean expressions is only four and the full diversity expression is simply $A_0+A_1$. 
However, implementation of Boolean algebra is rather tricky for large $M$. 
To address this, we adopt an alternative method for the algebraic function update realization. The truth table such as Karnaugh map can be used to evaluate the Boolean function. 
For every specific fading realization $\mathbf{A}=\mathbf{a}\in\{0,1\}^M$, we perform direct numerical evaluation of the corresponding CN and VN updates. By aggregating all the results, we can construct the truth tables of all messages and the output Boolean functions of DivE. 

For a fixed fading $\mathbf{A}=\mathbf{a}$, \emph{FadingMSD} performs the iterative Boolean updates over the protograph to determine the Boolean fading states of all messages and VN outputs as summarized in Algorithm~\ref{alg:FadingMSD}.
We denote the VN-to-CN and CN-to-VN message as $\alpha_{i\to j}^{(\ell)}\in\{0,1\}$ and $\beta_{j\to i}^{(\ell)}\in\{0,1\}$ at iteration $\ell$, respectively.
We also use ``$\cdot$'' and ``$+$'' to denote Boolean AND and OR, respectively.
By executing FadingMSD over all $2^M$ realizations of $\mathbf{a}$, we collect the binary outputs to construct the truth table of each VN, which can then be represented as the corresponding Boolean fading function in $\mathbf{A}$.

\begin{algorithm}[t]
\caption{FadingMSD($\cdot$)}
\label{alg:FadingMSD}
\begin{algorithmic}[1]
\footnotesize
\REQUIRE $\mathbb{G}_p,\mathbf{a}, {\pi(\cdot)}\in\{0,\ldots,M-1\}^n$
\ENSURE $\{\alpha_i^{(\ell_{\max})}\}_{i\in\mathcal{V}}$

\STATE \textit{// 1. Initialization Phase}
\STATE $\alpha_{i\to j}^{(0)} \gets a_{\pi(i)}$, $\forall (j,i)\in\mathcal{E}$

\STATE \textit{// 2. Main Iterative Loop}
\FOR{$\ell = 1,2,\ldots,\ell_{\max}$}
    \STATE \textit{// CN processing (AND)}
    \FOR{$j\in\mathcal{C}$}
        \FOR{$i\in\mathcal{N}_c(j)$}
            \STATE $\beta_{j\to i}^{(\ell)} \gets \displaystyle\prod_{i'\in\mathcal{N}_c(j)\setminus\{i\}} \alpha_{i'\to j}^{(\ell-1)}$
        \ENDFOR
    \ENDFOR

    \STATE \textit{// VN processing (OR)}
    \FOR{$i\in\mathcal{V}$}
        \FOR{$j\in\mathcal{N}_v(i)$}
            \STATE $\alpha_{i\to j}^{(\ell)} \gets \displaystyle a_{\pi(i)} \;+\!\!\!\sum_{j'\in\mathcal{N}_v(i)\setminus\{j\}} \beta_{j'\to i}^{(\ell)}$
        \ENDFOR
    \ENDFOR

    \STATE \textit{// 3. Update Boolean fading state of all VNs}
    \FOR{$i\in\mathcal{V}$}
        \STATE $\alpha_i^{(\ell)} \gets \displaystyle a_{\pi(i)} \;+\!\!\!\sum_{j'\in\mathcal{N}_v(i)} \beta_{j'\to i}^{(\ell)}$
    \ENDFOR
\ENDFOR

\STATE \textbf{return} $\{\alpha_i^{(\ell_{\max})}\}_{i\in\mathcal{V}}$
\end{algorithmic}
\end{algorithm}

\subsection{Analysis of DivE}
Using DivE to evaluate the VN diversity order at each iteration, we observe that the number of VNs that satisfy full diversity increases as decoding iteration proceeds. 
This increase is induced by a generalized rootcheck structure that arises during iterative decoding, which we call a \emph{generalized rootcheck}.

\begin{definition}[Generalized rootcheck]
\label{def:gen_rootcheck}
Under a block mapping $\pi(\cdot)$, CN $c_j$ is a \emph{generalized rootcheck} at iteration $\ell$ if there exists $m\neq \pi(i)$ such that

$
F_{v_{i'}\to c_j}^{(\ell)} \in \left\{\, A_m,\; A_0+\cdots+A_{M-1} \,\right\},
 \forall\, i' \in \mathcal{N}_c(j)\setminus\{i\}.
$
\end{definition}

\begin{figure}[t]
\centering
\begin{minipage}{0.2\textwidth}
    \centering
    \includegraphics[width=0.75\textwidth]{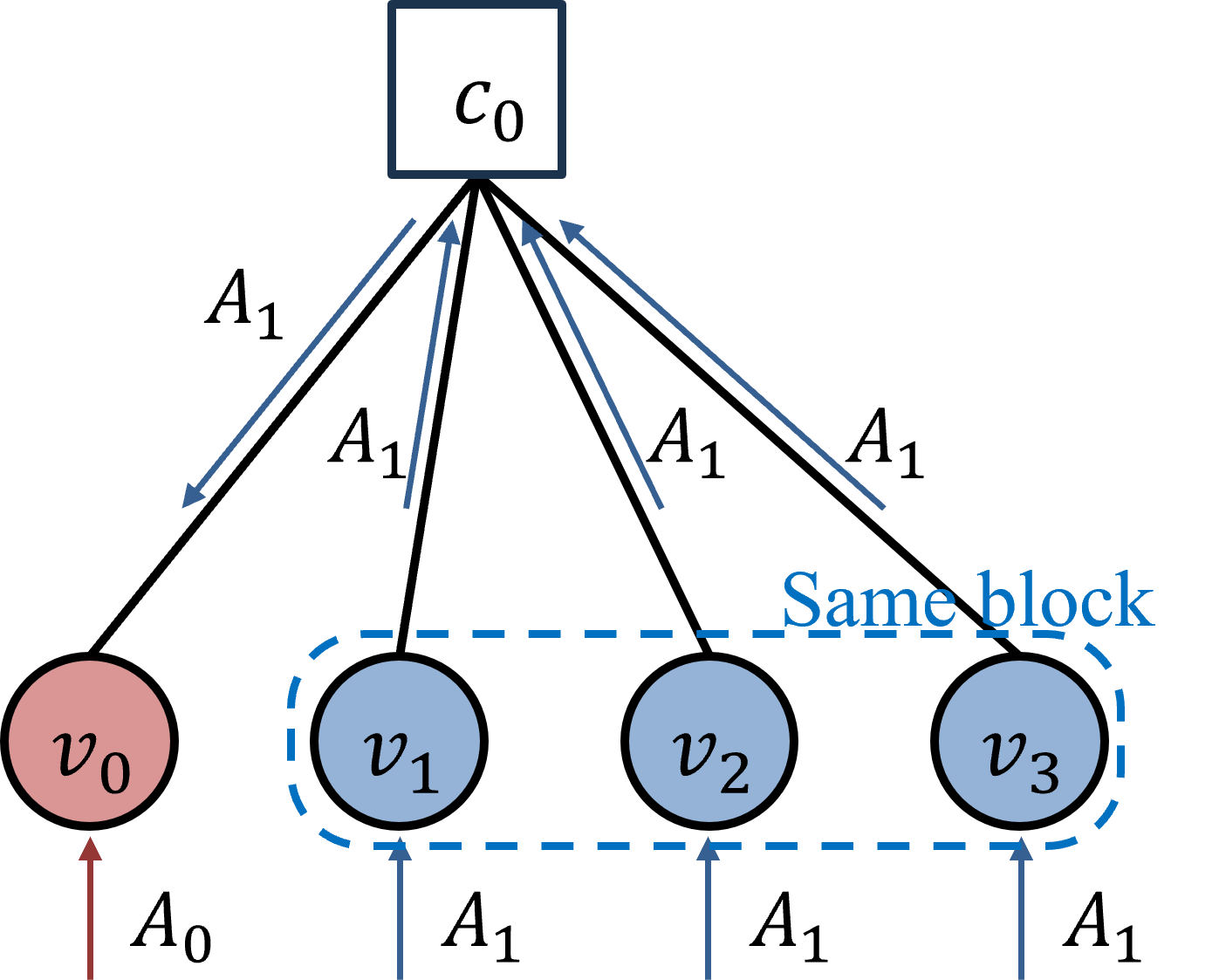}
    \subcaption{}
    \label{fig:con_rootcheck}
\end{minipage}
\hfill
\begin{minipage}{0.28\textwidth}
    \centering
    \includegraphics[width=0.75\textwidth]{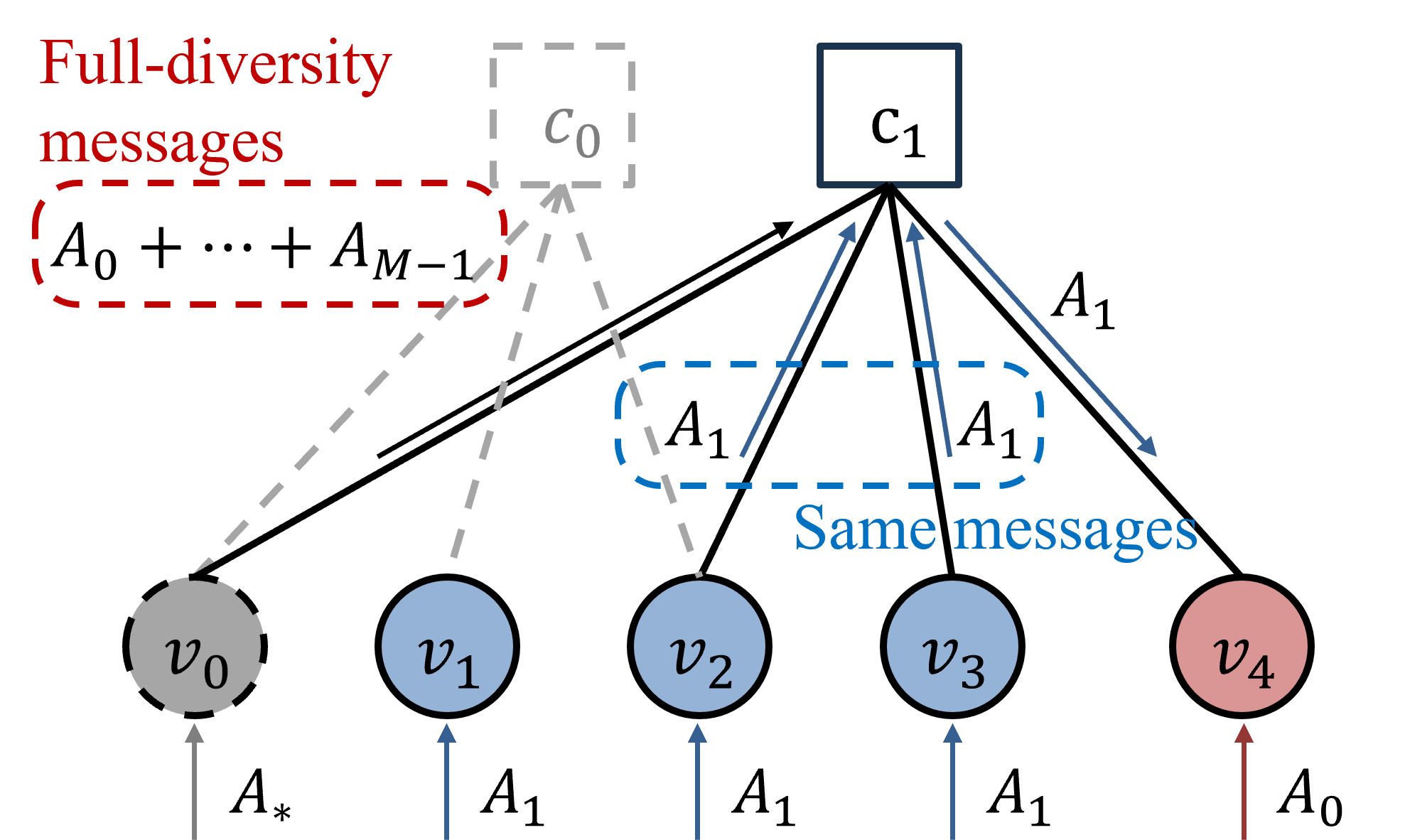}
    \subcaption{}
    \label{fig:gen_rootcheck}
\end{minipage}
\caption{Structure of (a) rootcheck and (b) generalized rootcheck: Rootcheck is block-mapping-based whereas generalized rootcheck is message-based. }
\label{fig:rootcheck_structure}
\end{figure}

This structure is illustrated in Fig.~\ref{fig:rootcheck_structure}. 
Each VN is colored according to its assigned fading block, and the labels on the edges denote the Boolean fading functions propagated along the corresponding edges.
A conventional rootcheck is determined by the block indices assigned to the VNs connected to a CN, as shown in Fig.~\ref{fig:con_rootcheck}.
In contrast, a generalized rootcheck is induced by the incoming messages: when all other incoming VN-to-CN messages are either full diversity messages or the same single-block fading function ($A_m$), as shown in Fig.~\ref{fig:gen_rootcheck}. Once $v_0$ becomes full diversity at iteration $\ell-1$ and sends $A_0 + \cdots + A_{M-1}$ to $c_1$ forms a generalized rootcheck at iteration $\ell$.
For both a conventional rootcheck and a generalized rootcheck, the CN update delivers the same single-block Boolean fading function (e.g., $A_1$) to the target VN.
Therefore, generalized rootchecks include conventional rootchecks, and they can be formed as iteration proceeds.

Since the set of full diversity VNs is determined by the block mapping through the generalized rootchecks, we can search for a suitable block mapping to attain the optimal coded diversity.
In particular, our objective is to align full diversity with the information VNs; we refer to this as \emph{diversity alignment (DA)}.

\begin{algorithm}[t]
\caption{DA block-mapping search for $M=2$ (overview)}
\label{alg:DA_overview}
\begin{algorithmic}[1]
\footnotesize
\REQUIRE Protograph $\mathbb{G}_p$, maximum trials $T$
\ENSURE A DA block mapping $\pi$ or \textbf{FAIL}

\STATE Start from $R=1/2$
\WHILE{no feasible mapping is found}
    \STATE Generate candidate partial mappings by \textbf{Pre-assignment~1--3}.
    \FOR{each candidate partial mapping}
        \FOR{$t=1$ to $T$}
            \STATE Greedy assign the remaining VNs to generate generalized rootcheck.
            \IF{All info. VNs have full diversity (using DivE)}
                \STATE \textbf{return} $\pi$
            \ENDIF
        \ENDFOR
    \ENDFOR
    \STATE Reduce the rate by including additional parity VNs and repeat.
\ENDWHILE

\STATE \textbf{return} \textbf{FAIL}
\end{algorithmic}
\end{algorithm}

%%%%%%%%%%%%%%%%%%%%%%%%%%%%%%%%%%%%%%%%%%%%%%%%%%%%%%%%%%%%%%%%%%%
%Search algorithm
\section{A Block Mapping Search Algorithm for Diversity Alignment}
\label{sec:DA_blockmapping}
This section presents a greedy search algorithm to find a DA block mapping for a given LDPC protograph over a BFC with $M=2$.
Our goal is to find a block mapping such that all information VNs achieve full diversity under iterative decoding.
To achieve this, we use DivE to track the full diversity VNs and incrementally assign the remaining VNs to promote generalized rootcheck configurations.

We first apply a pre-assignment to prevent parity VNs from achieving full diversity and reduce the search space.
After pre-assignment, greedily complete the remaining unassigned VNs.
If full diversity is not feasible at the nominal AWGNC-optimized rate, we progressively include additional parity VNs and search for the maximum feasible rate.
The overall search procedure operates as summarized in Algorithm~\ref{alg:DA_overview}.

\subsection{Pre-assignment}
\label{subsec:pre_assignment}

\subsubsection*{Pre-assignment~1 (Stopping set prevention)}
As shown in Fig.~\ref{fig:Init1} structure, if two VNs with the same neighboring CN set are assigned to the same block, neither VN achieves full diversity.
To avoid this, we assign such VNs to different blocks.

\subsubsection*{Pre-assignment~2 (Common block assignment for parity VNs)}
As shown in Fig.~\ref{fig:Init2}, assigning the parity VNs to the same block so that the DA block mapping search can assign the information VNs in the other block and form generalized rootchecks that yield full diversity.
We first assign blocks to the four VNs in the dual-diagonal parity part of Fig.~\ref{fig:BG1}.
Specifically, we enumerate the block mappings
$(\pi(p_0),\pi(p_1),\pi(p_2),\pi(p_3)) \in \{0,1\}^4$, while avoiding redundant cases obtained by swapping the two blocks.
For each CN, we assign any unassigned neighboring parity VNs to the same block as the already assigned parity neighbors.

\subsubsection*{Pre-assignment~3 (Handling punctured information VNs)}
\begin{figure}[t]
\centering
\begin{subfigure}[t]{0.25\linewidth}
  \centering
  \includegraphics[width=\linewidth]{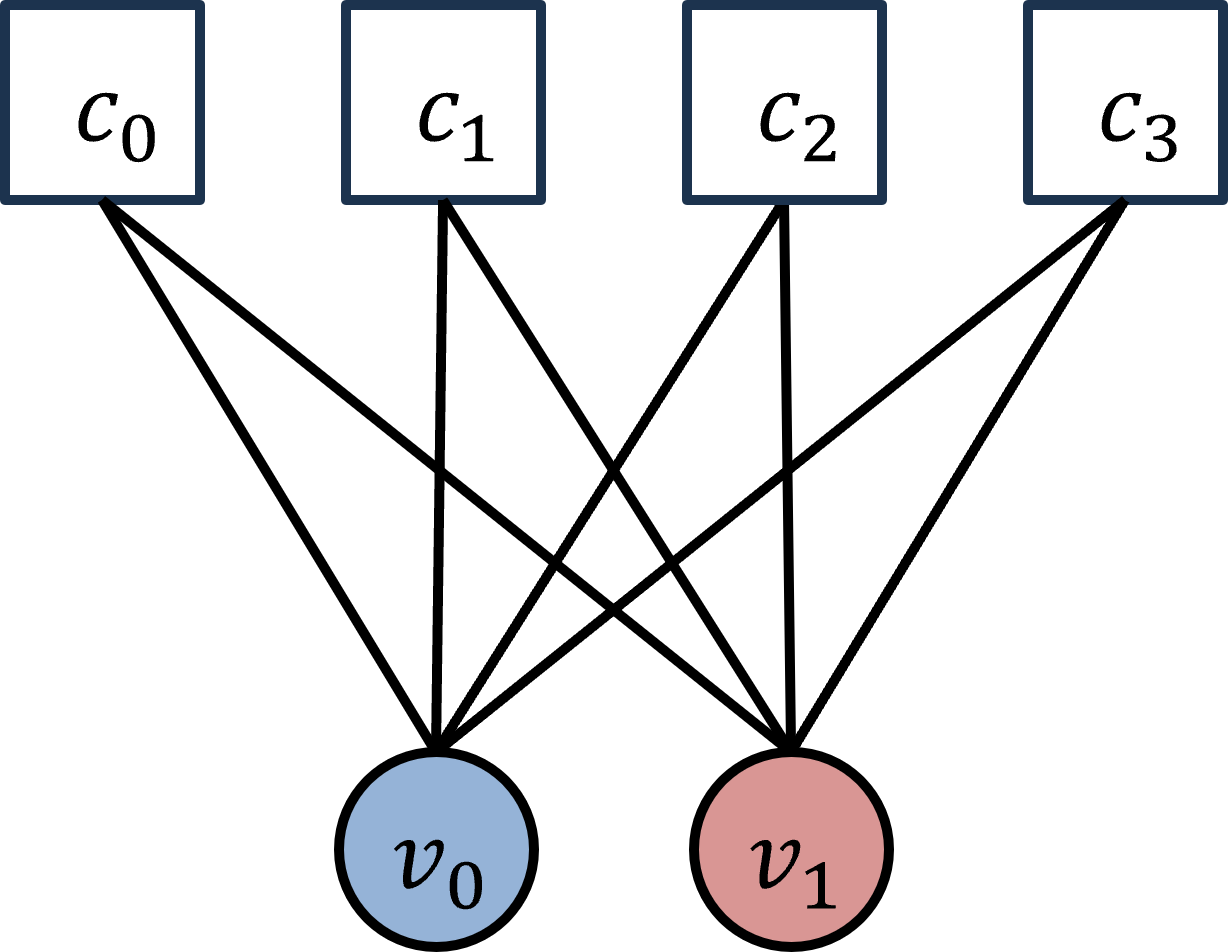}
  \subcaption{}
  \label{fig:Init1}
\end{subfigure}
\begin{subfigure}[t]{0.32\linewidth}
  \centering
  \includegraphics[width=\linewidth]{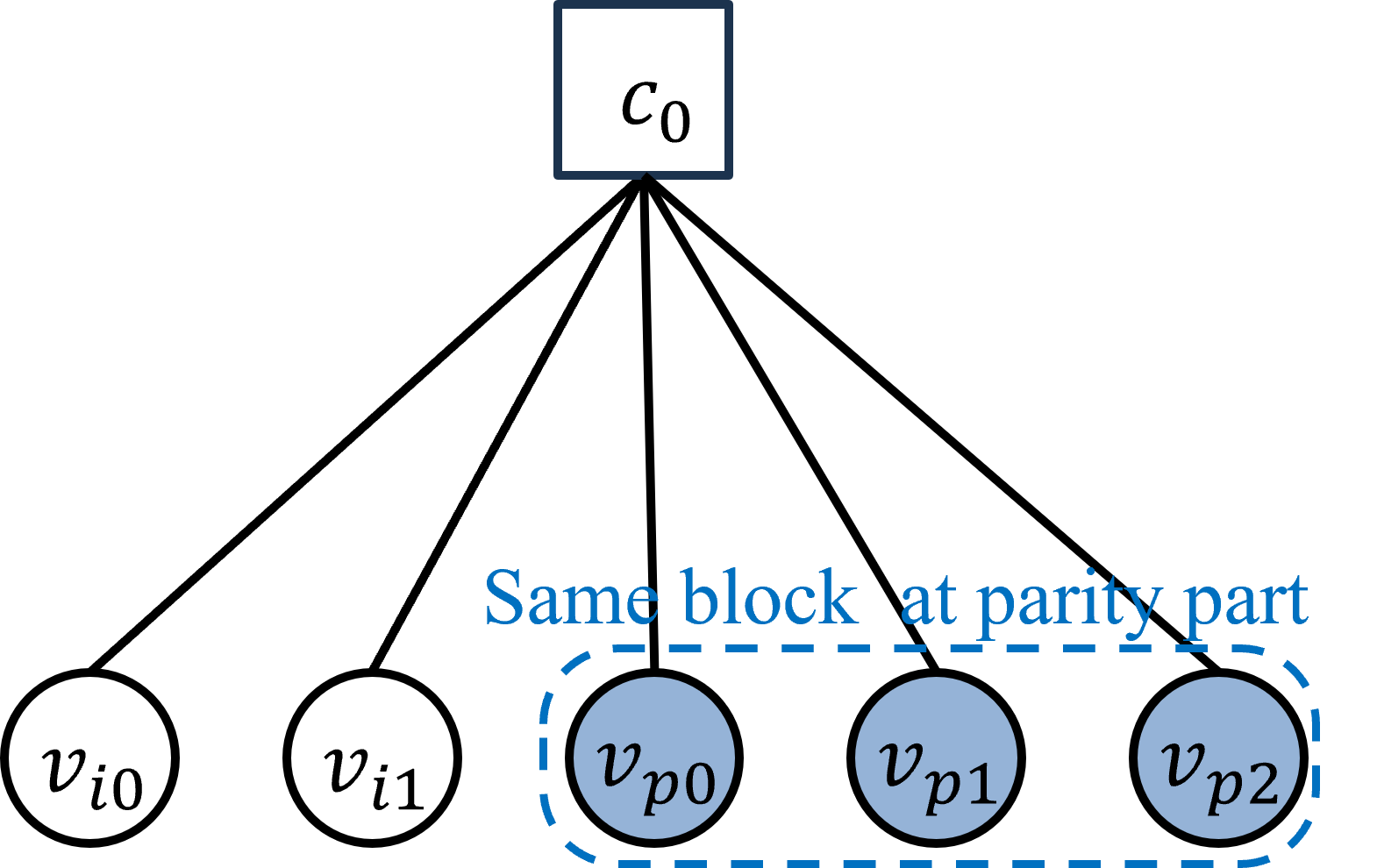}
  \subcaption{}
  \label{fig:Init2}
\end{subfigure}
\begin{subfigure}[t]{0.33\linewidth}
  \centering
  \includegraphics[width=\linewidth]{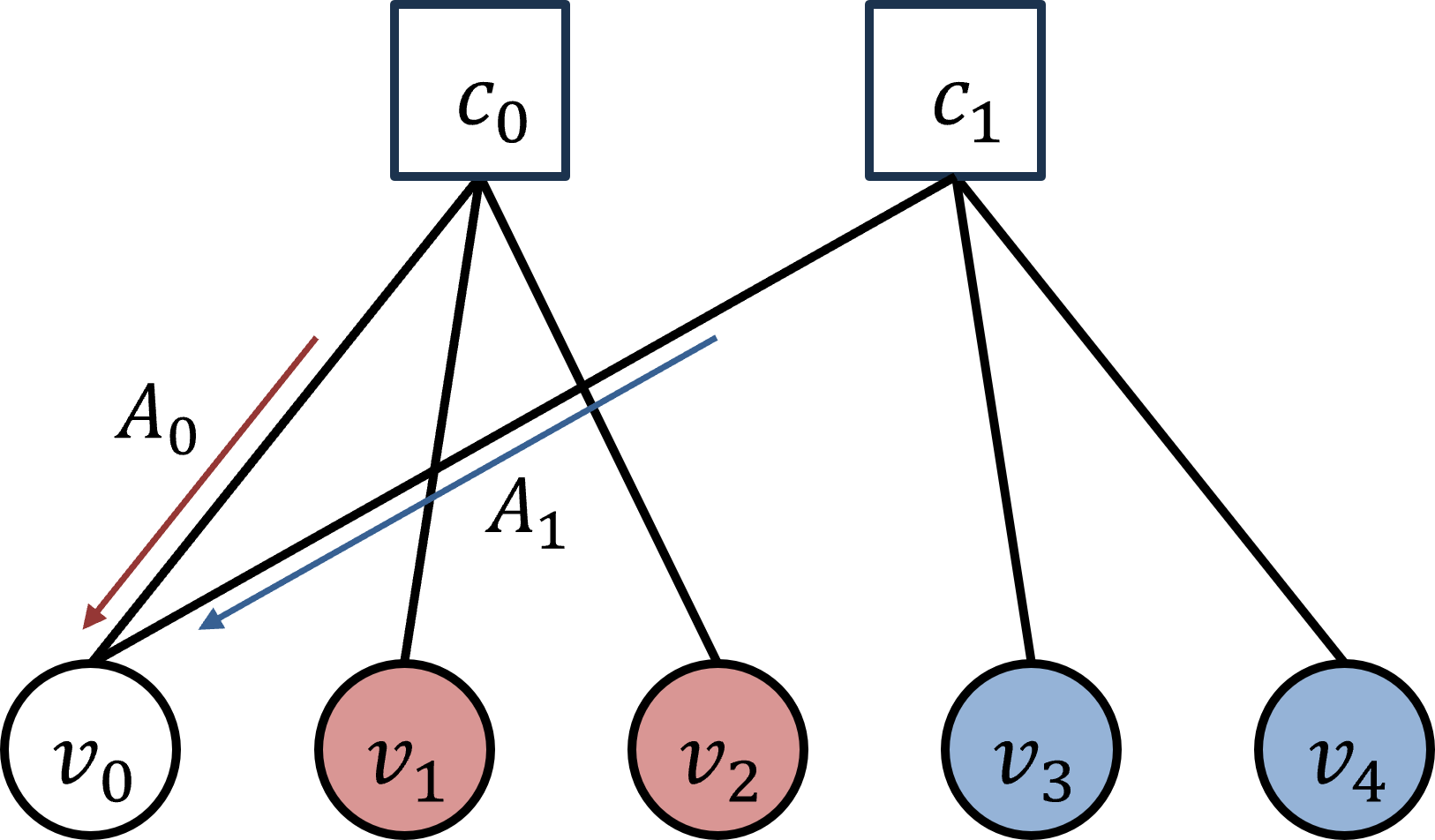}
  \subcaption{}
  \label{fig:Init3}
\end{subfigure}

\caption{Pre-assignment steps: (a) stopping set prevention, (b) common block assignment for parity VNs, (c) handling punctured information VNs (for information puncturing of 5G-NR).  }
\label{fig:xxx_abc}
\end{figure}

Pre-assignment~3 handles punctured information VNs in the 5G-NR BGs.
Since punctured VNs are not transmitted by fading blocks, they should receive each Boolean fading function $A_m$ through incoming CN messages.
As shown in Fig.~\ref{fig:Init3}, each punctured VN selects two adjacent CNs, so that the punctured VN receives the $A_0$ and $A_1$ Boolean fading functions in the two-block case.

\subsection{Generalized Rootcheck-based Search}
Generalized rootcheck-based search algorithm's objective is to complete the remaining block assignments so that the resulting mapping can induce generalized rootcheck structures under the DivE criteria, thereby enabling full diversity for all information VNs.
Starting from each candidate partial mapping produced by pre-assignment~1--3, search algorithm performs a greedy completion with randomized choices to avoid overly restrictive deterministic decisions.

\noindent \emph{Greedy search rule:}
Let $u(j)$ denote the number of unassigned neighboring VNs adjacent to CN $c_j$.
At each step, CNs are grouped according to $u(j)$ and examined in ascending order.
For a selected CN group, we use DivE to test whether there exists a block assignment of its neighboring VNs that can yield a generalized rootcheck structure.
If such an assignment exists, we apply it to ensure that at least one adjacent information VN attains full diversity.
If feasible local assignments exist, one is selected at random and applied; otherwise, the search moves to the next CN group with a larger $u(j)$.
If no further assignment is possible or if the resulting block mapping fails to achieve full diversity for all information VNs, the current attempt is declared fail and the same procedure is repeated for the next pre-assignment candidate.

%%%%%%%%%%%%%%%%%%%%%%%%%%%%%%%%%%%%%%%%%%%%%%%%%%%%%%%%%%%%%%%%%%%
\section{Numerical Results}
\label{sec:num_results}

We evaluate the proposed DivE-based block-mapping search on the standardized 5G-NR LDPC base graphs BG1 and BG2 over a two-block BFC with BPSK modulation.
For BG1, we use the lifting size $Z=240$ $((N,R)=(11040,22/46))$, whereas for BG2 we use $Z=20$, $((N,R)=(480,10/24))$.
Since $R=1/2$ is diversity-optimal for the two-block case, we first search for a valid mapping at $R=1/2$; if unsuccessful, we progressively include parity VNs and repeat the search
We obtain multiple full diversity mappings at $R=22/46(\approx0.478)$ for BG1 and $R=10/24(\approx0.417)$ for BG2.
These code rates are lower than the optimal rate of $1/2$, we attribute the observed shortcomings to the high rate precoding structure in 5G-NR LDPC BGs, whose detailed investigation is omitted due to page limitations.
The block mappings for BG1 and BG2 obtained by the proposed search algorithm are summarized in Tables~\ref{tab:BG1_block_mapping} and~\ref{tab:BG2_block_mapping}, respectively.
In contrast, for random block mapping, full diversity for all information VNs is achieved only at lower code rates, i.e., $R=22/54(\approx0.407)$ for BG1 and $R=10/26(\approx0.384)$ for BG2, demonstrating that the proposed block mapping enables full diversity operation at higher code rates.

\begin{table}[t]
\centering
\caption{BG1 Block mapping for $M=2$ ($R=22/46$).}
\label{tab:BG1_block_mapping}
\renewcommand{\arraystretch}{1.0}
\setlength{\tabcolsep}{4pt}
\scriptsize

\resizebox{\columnwidth}{!}{%
\begin{tabular}{@{}c l@{}}
\toprule
Block & Assigned VN indices \\ \midrule
1 & $\mathcal{V}_1=\{0,8,10,14,15,16,18,20,21,22,24,25,26,28,31,32,33,36,37,40,41,43,44\}$ \\
2 & $\mathcal{V}_2=\{1,2,3,4,5,6,7,9,11,12,13,17,19,23,27,29,30,34,35,38,39,42,45\}$ \\
\bottomrule
\end{tabular}%
}
\end{table}

\begin{figure}[t]
    \centering
    \includegraphics[width=0.9\linewidth]{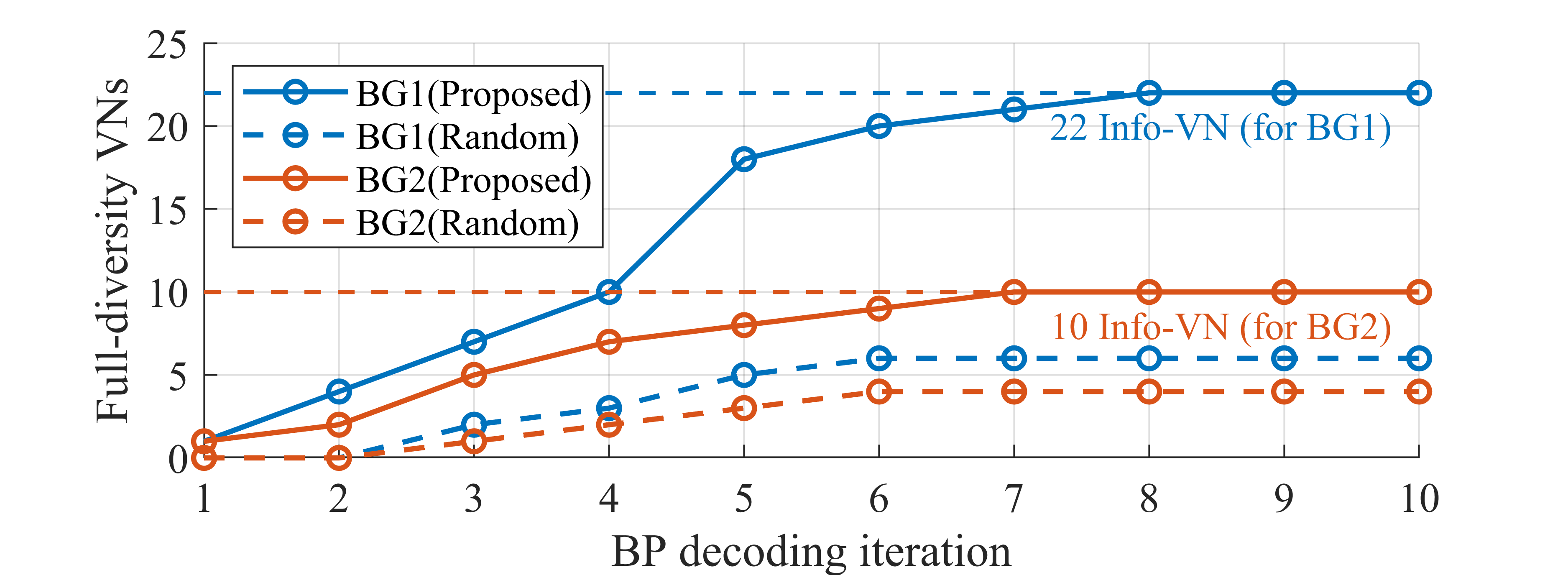}
    \caption{Number of full diversity info-VNs at each BP decoding iteration ($R\approx0.478$ for BG1, $R\approx0.417$ for BG2).}
    \label{fig_iterative}
\end{figure}

\begin{table}[t] 
\centering
\caption{BG2 Block mapping for $M=2$ ($R=10/24$).}
\label{tab:BG2_block_mapping}
\renewcommand{\arraystretch}{1.1}
\setlength{\tabcolsep}{5pt}
\small
\begin{tabular}{@{}c p{0.78\columnwidth}@{}}
\toprule
Block & Assigned VN indices \\ \midrule
1 & $\mathcal{V}_1=\{0,1,3,8,9,10,12,13,14,17,20,23\}$ \\
2 & $\mathcal{V}_2=\{2,4,5,6,7,11,15,16,18,19,21,22\}$ \\
\bottomrule
\end{tabular}
\end{table}

Fig.~\ref{fig_iterative} shows the number of information VNs satisfying full diversity at each BP decoding iteration, evaluated by the DivE.
For both BG1 and BG2, the number of full diversity information VNs progressively increases through iterative decoding.
For the block mapping found by the proposed algorithm, for BG1, all $22$ information VNs satisfy full diversity at the $8$th iteration, and for BG2, all $10$ information VNs satisfy full diversity at the $7$th iteration.
In contrast, when random block mapping is applied at the same code rate BGs, only $6$ information VNs in BG1 and $4$ information VNs in BG2 satisfy full diversity.

\begin{figure}[t]
    \centering
    \includegraphics[width=0.9\linewidth]{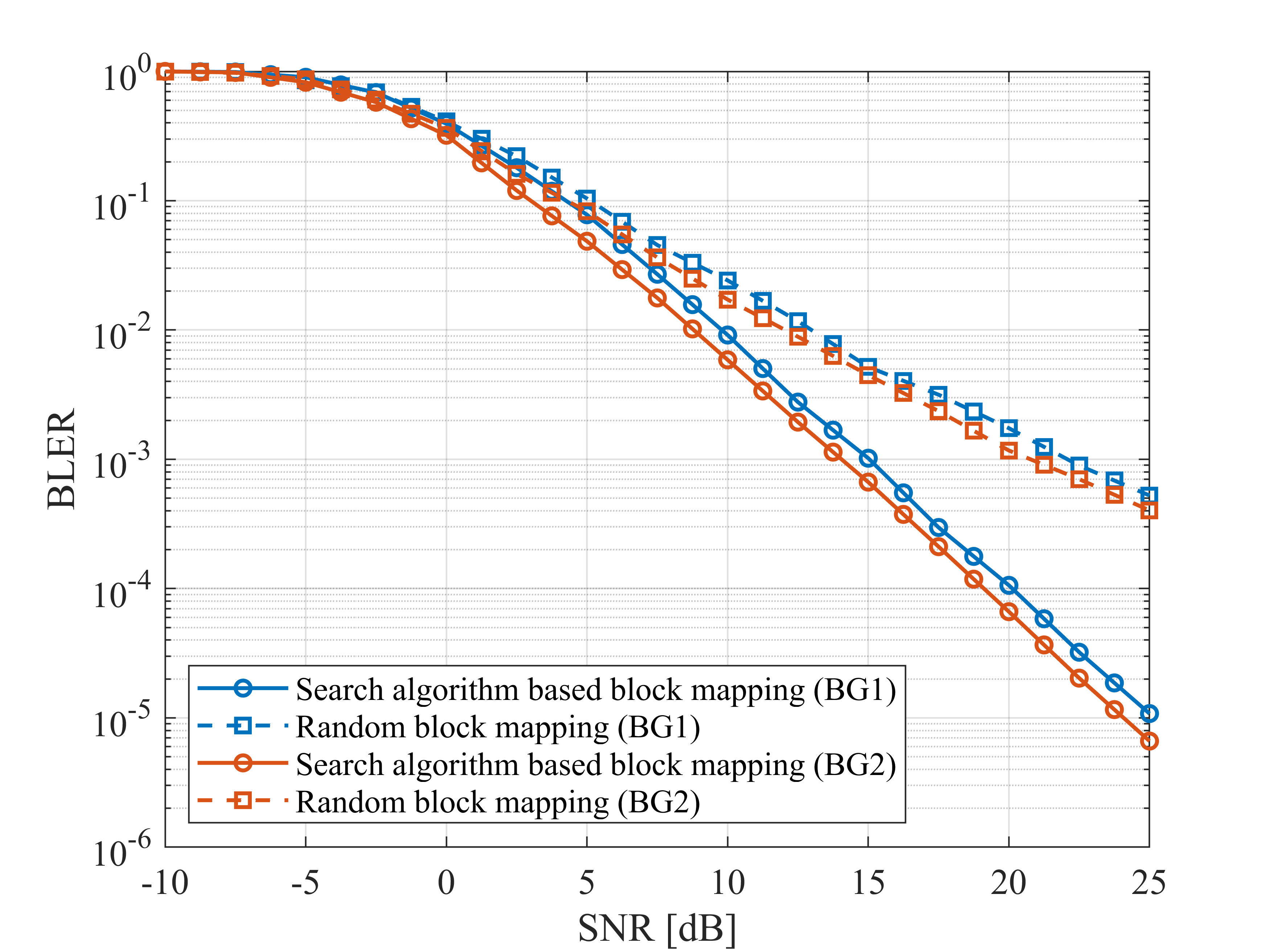}
    \caption{BLER of 5G-NR LDPC BGs with proposed block mapping and random mapping, BG1 $(N, R)=(11040, 22/46)$, BG2 $(N, R)=(480, 10/24)$, BPSK.}
    \label{fig_bg_bpsk}
    \vspace{-0.1in}
\end{figure}

From Fig.~\ref{fig_bg_bpsk}, the proposed block mappings achieve full diversity under Rayleigh BFCs with $M=2$, whereas random block mappings are diversity-limited, resulting in a smaller high-SNR slope.
Consequently, for both base graphs the proposed mappings achieve a steeper high-SNR slope and a substantially lower BLER than random mappings.

%%%%%%%%%%%%%%%%%%%%%%%%%%%%%%%%%%%%%%%%%%%%%%%%%%%%%%%%%%%%%%%%%%%
\section{Conclusion}
\label{sec:conclusion}

In this work, we proposed a novel framework to enable full diversity for standardized 5G-NR LDPC codes over block-fading channels. We introduced Diversity Evolution (DivE), which leverages Boolean functional updates to analyze the exact diversity behavior of variable nodes under iterative decoding.
This analysis led to the identification of generalized rootchecks—dynamic structures that can be analyzed in an iteration-wise manner and provide full diversity gains without explicit rootcheck constraints in the base matrix.

Leveraging these insights, we developed a greedy block-mapping search that systematically aligns information bits with full diversity. Our results on 5G-NR base graphs demonstrate that the proposed mappings achieve full diversity and significantly outperform random mappings. These findings substantiate that standard 5G-NR codes possess the inherent capability to operate as Root-LDPC codes in a broader sense, provided that the block mapping is properly optimized.

For future work, we can study block-mapping design for BFCs with more than two blocks ($M>2$) and develop mapping strategies that achieve the optimal diversity order in this general setting.

\end{document}